\documentclass[preprint,aps,floatfix,onecolumn]{revtex4}
\usepackage{epsfig,amsmath,bbm,graphicx,color}

\begin{document}

\title{Dependence of protein mechanical unfolding pathways on pulling speeds}

\author{\bf Mai Suan Li and Maksim Kouza}

\affiliation{Institute of Physics, Polish Academy of Sciences,
Aleja Lotnikow 32/46, 02-668 Warszawa, Poland}

\begin{abstract}
Mechanical unfolding of the fourth domain of
 {\em Distyostelium discoideum} filamin (DDFLN4) was studied in detail
using the C$_{\alpha}$-Go model.
We show that
unfolding pathways of this protein depend on the pulling speed.
The agreement between theoretical and experimental results
on the sequencing of unfolding events is achieved at low loading rates.
The unfolding free energy landscape is also constructed using dependencies
of unfolding forces on pulling speeds.
\end{abstract}

\maketitle

\vfill

\section{Introduction}

The last ten years have witnessed an intense activity in single-molecule
force spectroscopy experiments in detecting
inter and
intramolecular forces of biological systems to understand
their functions and structures.
Much of the research has been focused
on elastic properties of proteins, DNA, and RNA, i.e, their response to an external force, following the seminal papers by Rief {\em et al.} \cite{Rief_Science97}, and Tskhovrebova {\em et al.} \cite{Tskhovrebova_Nature97}.
The main advantage of this technique is its ability
to separate out the fluctuations of individual trajectories from the
ensemble average behavior observed in  traditional bulk biochemical
experiments. This allows for studying unfolding
pathways in detail using the end-to-end distance as a reaction
coordinate. Moreover, the single-molecule
force spectroscopy can be used to decipher the unfolding
free energy landscape (FEL) of biomolecules \cite{Bustamante_ARBiochem_04,MSLi_JCP08}.

As cytoskeletal proteins, large actin-binding proteins play a key roles
in cell organization, mechanics and signalling\cite{Stossel_NRMCB01}. 
During the process of permanent cytoskeleton reorganization, all
involved participants are subject to mechanical stress. One
of them is the fourth domain
{\em Distyostelium discoideum} filamin (DDFLN4),
which binds different components of
actin-binding protein. Therefore, understanding the  mechanical response of
this domain to a stretched force is of great interest.
Recently, using the AFM experiments, 
Schwaiger {\em et al.}  \cite{Schwaiger_NSMB04,Schwaiger_EMBO05} have obtained two major results for DDFLN4.
First,  this domain (Fig. \ref{native_ddfln4_strands_fig})
 unfolds via intermediates as the force-extension curve displays two peaks
centered at the end-to-end extension $\Delta R \approx 12$ nm and
$\Delta R \approx 22$ nm.
Second, with the help of  loop mutations, it was suggested
that during the first unfolding event (first peak) strands A and  B
unfold first.
Therefore, strands C - G form a stable intermediate structure, which then
unfolds in the second unfolding event (second peak).
In addition, Schwaiger {\em et al.}
\cite{Schwaiger_EMBO05} have also determined the
FEL parameters of DDFLN4. 

With the help of the C$_{\alpha}$-Go model \cite{Clementi_JMB00}, Li {\em et al.}
\cite{MSLi_JCP08}
have demonstrated that the mechanical unfolding of DDFLN4 does follow
 the three-state
scenario but the full agreement between theory and experiments was not
obtained. The simulations \cite{MSLi_JCP08} showed
that two peaks in the force-extension profile occur
at $\Delta R \approx 1.5$ nm and 11 nm, i.e.,
the Go modeling does not detect the peak
at $\Delta R \approx 22$ nm. Instead, it predicts the existence of
a peak not far from the native
conformation. More importantly, theoretical unfolding pathways
\cite{MSLi_JCP08} are very different from the
experimental ones \cite{Schwaiger_NSMB04}:
the unfolding initiates from the C-terminal,
but not from the N-terminal terminal as shown by the experiments.

It should be noted that the pulling speed used in the previous simulations
is about five orders of magnitude larger than
the experimental value \cite{Schwaiger_NSMB04}.
Therefore, a natural
question
emerges is if the discrepancy between theory and experiments is due
to huge difference in pulling speeds.
Motivated by this, we have carried low-$v$ simulations, using the Go
model \cite{Clementi_JMB00}.
Interestingly,
we uncovered that unfolding pathways of DDFLN4 depend on the pulling speed and
only at
$v \sim 10^4$ nm/s, the theoretical unfolding sequencing coincides with
the experimental one \cite{Schwaiger_NSMB04}.
However, even at low loading rates,
the existence of the peak at $\Delta R \approx 1.5$ nm
remains robust 
and the Go modeling does not capture the maximum at $\Delta R \approx 22$ nm.

In the previous work \cite{MSLi_JCP08},
using dependencies
of unfolding times on external forces,
 the distance between the native state (NS) and intermediate state
(IS) $x_{u1}$, and
the distance between the IS and denaturated state (DS) $x_{u2}$
of DDFLN4
have been estimated.
In the Bell approximation, the agreement between the theory and
experiments \cite{Schwaiger_EMBO05} was reasonable.
However, in the non-Bell approximation
\cite{Dudko_PRL06}, the theoretical values of  $x_{u1}$, and $x_{u2}$
seem to be high \cite{MSLi_JCP08}.
In addition the unfolding barrier between the first transition state
(TS1) and NS
$\Delta G^{\ddagger}_1$ is clearly higher than its experimental
counterpart (Table 1).

In this paper, assuming that the microscopic kinetic theory
\cite{Dudko_PRL06} holds for a three-state protein, we calculated
$x_{ui} (i=1,2)$ and unfolding barriers
by a different method
which is based on dependencies of peaks in the force-extension curve
on $v$. Our present estimations for
the unfolding FEL parameters are more reasonable
compared to the previous ones \cite{MSLi_JCP08}.
Finally, we have also studied thermal unfolding 
pathways of DDFLN4 and shown
that the
mechanical unfolding pathways are different from the thermal ones.

\section{Method}

\begin{figure}
\epsfxsize=6.3in
\vspace{0.2in}
\centerline{\epsffile{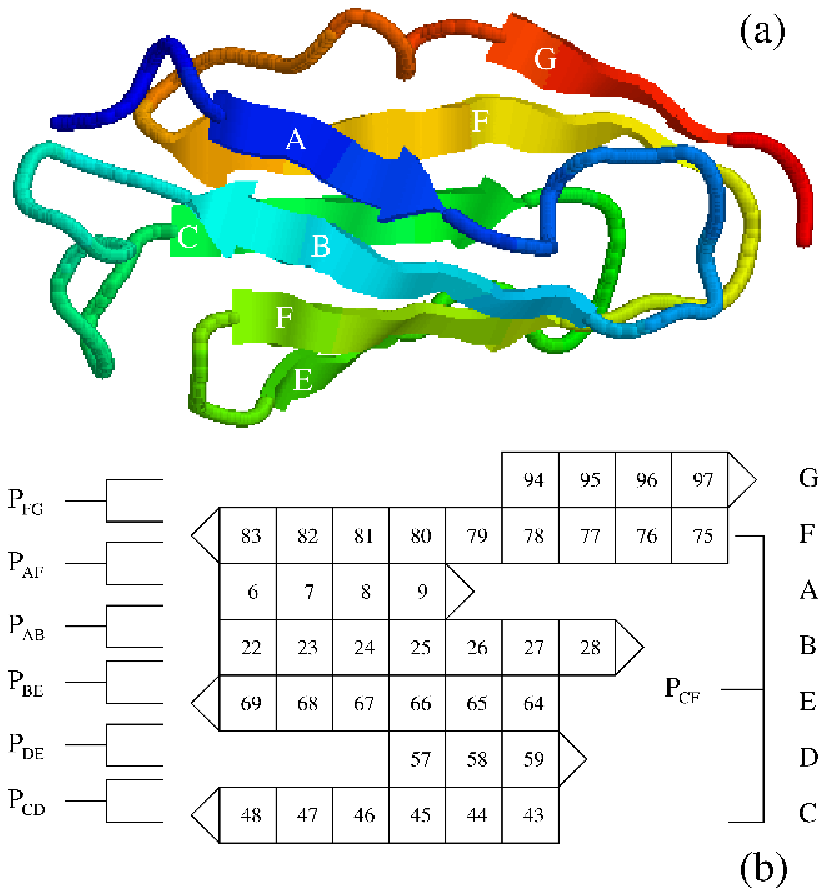}}
\caption{}
\label{native_ddfln4_strands_fig}
\end{figure}

The native conformation
of DDFLN4, which has seven $\beta$-strands, enumerated as
A to G,
was taken from the PDB (PI: 1KSR,
Fig. \ref{native_ddfln4_strands_fig}a).
We assume that
residues $i$ and $j$ are in native contact if the distance
 between them in the native conformation,
is shorter than a cutoff distance $d_c =
6.5$ \AA.
With this choice of $d_c$, the molecule has 163 native contacts.
Native contacts exist between seven pairs
of $\beta$-strands
P$_{\textrm{AB}}$,
P$_{\textrm{AF}}$, P$_{\textrm{BE}}$,
P$_{\textrm{CD}}$, P$_{\textrm{CF}}$, P$_{\textrm{DE}}$, and P$_{\textrm{FG}}$
(Fig. \ref{native_ddfln4_strands_fig}b).

We used the C$_{\alpha}$-Go model \cite{Clementi_JMB00} for a molecule.
The corresponding parameters of this model are chosen as follows
\cite{MSLi_BJ07,Kouza_JCP08}:
 $K_r =
100 \epsilon _H/$\AA$^2$, $K_{\theta} = 20 \epsilon _H/$rad$^2$,
$ K_{\phi}^{(1)} = \epsilon _H$, and
$K_{\phi}^{(3)} = 0.5 \epsilon _H$, where $\epsilon_H$ is the
characteristic hydrogen bond energy, and $C = 4$ \AA.
As in our previous works  \cite{MSLi_BJ07,Kouza_JCP08}, we set
$\epsilon _H = 0.98 \, {\rm kcal/mol}$. Then, the temperature $T=285$ K
corresponds to $0.53 \epsilon _H/k_B$ and all
computations have been 
performed at this temperature.
The force
unit is $[f] = \epsilon_H/$\AA $\, = 68$ pN \cite{MSLi_BJ07}.

The simulations were carried out in the over-damped limit
with the water viscosity $\zeta = 50\frac{m}{\tau_L}$
\cite{Veitshans_FD97}, where 
the time unit $\tau_L = (ma^2/\epsilon_H)^{1/2} \approx 3$ ps,
$m$ is a typical mass of amino-acids, and $a=4 \AA \,$ a distance between
two neighboring residues.
Neglecting  the inertia term, the Brownian dynamics
  equation was numerically solved by the simple Euler method.
Due to the large viscosity, we can choose a large time step
$\Delta t = 0.1 \tau_L$, and this choice allows us to study unfolding at low
loading rates.

In the constant velocity force simulations, we fix the N-terminal and pull the
C-terminal by applying the force $f = K_r(\nu t -r)$,
 where $r$ is the displacement of the pulled atom 
 from its original position
\cite{Lu_BJ98}, and the spring constant of cantilever $K_r$ is set
to be the same as  the spring constant of the Go model.
The pulling direction was chosen along the vector drawn
 from the fixed atom to the
pulled one.
                                                                                
The mechanical unfolding sequencing
was studied by
monitoring the fraction of native contacts of the $\beta$-strands
and of their seven pairs as a function of $\Delta R$,
which is admitted  a good reaction coordinate.
In order to probe thermal unfolding pathways, for the $i$-th trajectory
we introduce the progress variable $\delta _i =
t/\tau^i_{u}$, where $\tau^i_{u}$ is the unfolding time \cite{MSLi_BJ07}.
Then one
can average the fraction of native contacts over many trajectories
in a unique time window
$0 \le \delta _i \le 1$ and monitor the unfolding sequencing with
the  help of the progress variable $\delta$.

\section{Results}

\subsection{Robustness of peak at $\Delta R \approx 1.5$ nm and absence
of maximum at $\Delta R \approx 22$ nm at low pulling speeds}

In our previous high pulling speed
($v = 3.6\times 10^7$ nm/s) simulations
\cite{MSLi_JCP08}, the force-extension curve shows two
peaks at $\Delta R \approx 1.5$ nm and 10 nm, while the experiments
showed that peaks appear at $\Delta R \approx 12$ nm and 22 nm.
The question we ask if one can reproduce the experimental results at
low pulling speeds. Within our computational facilities, we were
able to perform simulations at the lowest $v = 2.6\times 10^4$ nm/s
which is about three orders of magnitude lower than that used
before \cite{MSLi_JCP08}.

\begin{figure}
\epsfxsize=6.3in
\vspace{0.2in}
\centerline{\epsffile{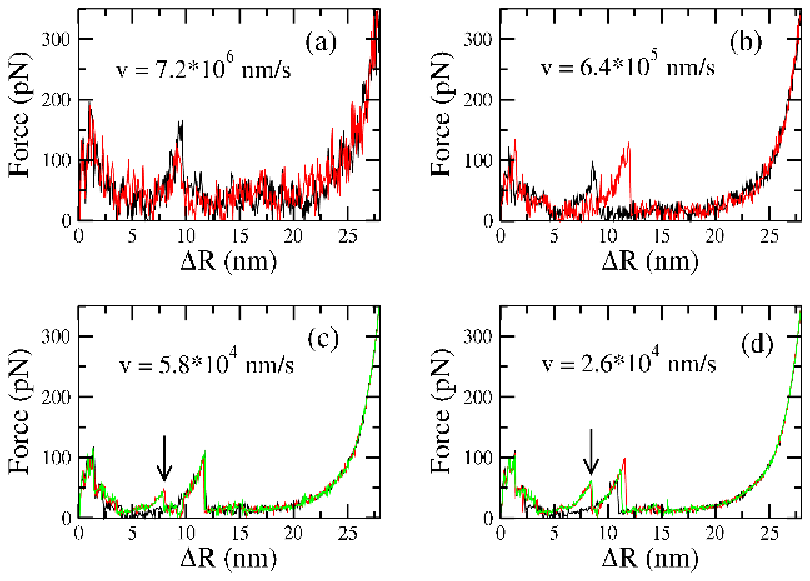}}
\caption{}
\label{force_ext_traj_fig}
\end{figure}

\begin{figure}
\epsfxsize=6.3in
\vspace{0.2in}
\centerline{\epsffile{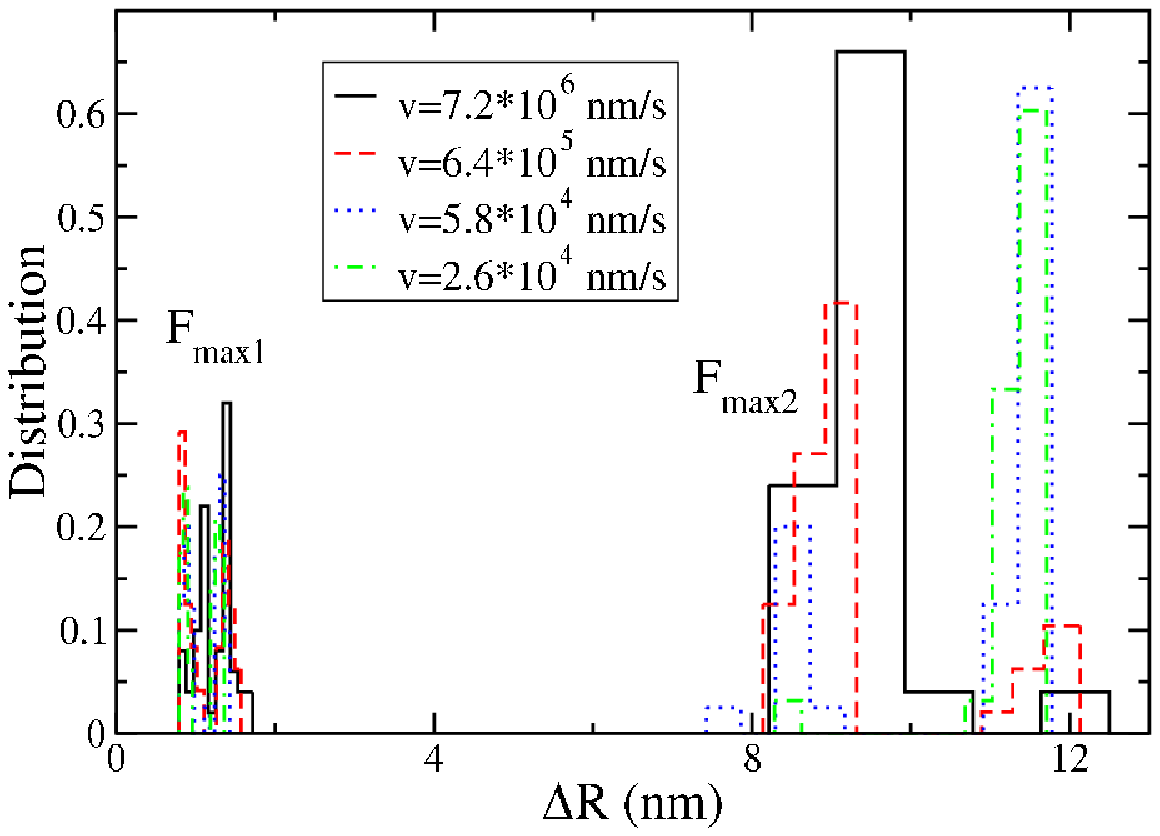}}
\caption{}
\label{dist_fmax_pos_fig}
\end{figure}

Fig. \ref{force_ext_traj_fig} show 
force-extension curves for four representative pulling speeds.
For the highest $v = 7.2\times 10^6$ nm/s
(Fig. \ref{force_ext_traj_fig}a), there are two peaks
located at extensions $\Delta R \approx 1.5$ nm and 9 nm.
As evident from Figs. \ref{force_ext_traj_fig}b, c and d,
the existence of the first peak remains robust against reduction of $v$.
Positions 
of $f_{max1}$ weakly fluctuate over the range
$0.9 \lesssim \Delta R \lesssim 1.8$ nm for all values of $v$
(Fig. \ref{dist_fmax_pos_fig}).
As $v$ is reduced, $f_{max1}$ decreases but this peak does not
vanish if one interpolate our results to the lowest pulling speed
$v_{exp} = 200$ nm/s
used in the experiments \cite{Schwaiger_NSMB04}
(see below). Thus, opposed to the experiments, the first peak
occurs already at small end-to-end extensions.
We do not exclude a possibility that such a peak was
overlooked  by experiments,
as it happened with the titin domain
I27. Recall that, for this domain the first
 AFM experiment \cite{Rief_Science97}
did not trace the hump which was observed in the later
simulations \cite{Lu_BJ98} and experiments \cite{Marszalek_Nature99}.

Positions of the second peak $f_{max2}$
are more scattered compared to $f_{max1}$, ranging
from about 8 nm to 12 nm (Fig. \ref{dist_fmax_pos_fig}). Overall, they
move toward higher values upon
reduction of $v$ (Fig. \ref{force_ext_traj_fig}). If at $v=6.4\times 10^5$ nm/s only about 15$\%$
trajectories display $\Delta R_{max2} > 10$ nm, then this percentage reaches
65$\%$ and 97\% for $v=5.8\times 10^4$ nm/s and $2.6\times 10^4$ nm/s,
respectively (Fig. \ref{dist_fmax_pos_fig}).

At low $v$, unfolding pathways show rich diversity. 
For $v \gtrsim 6.4\times 10^5$ nm/s, the force-extension profile shows
only two peaks in all trajectories
studied (Fig. \ref{force_ext_traj_fig}a and \ref{force_ext_traj_fig}b),while
for lower speeds $v = 5.8\times 10^4$ nm/s and $2.6\times 10^4$ nm/s,
about $4\%$ trajectories display even four peaks
(Fig. \ref{force_ext_traj_fig}c and 
\ref{force_ext_traj_fig}d), i.e. the four-state behavior.

We do not observe any peak at $\Delta R \approx 22$ nm for all 
loading rates (Fig. \ref{force_ext_traj_fig}),
and it is very unlikely that it will appear at lower
values of $v$.
Thus, the Go model, in which non-native interactions are neglected,
fails to reproduce this experimental observation.
Whether inclusion of non-native interactions would
cure this problem requires further studies.

\subsection{Dependence of mechanical unfolding pathways on loading rates}

The considerable fluctuations of peak positions and
occurrence of even three peaks already suggest that unfolding
pathways, which are kinetic in nature, may change if $v$ is varied.
To clarify this point in more detail, we show $\Delta R$-dependencies
of native contacts of all $\beta$-strands
and their pairs for $v=7.2\times 10^6$ nm/s
(Fig. \ref{cont_ext_v6_fig}) and $v=2.6\times 10^4$ nm/s 
(Fig. \ref{cont_ext_v15_fig}). For $v=7.2\times 10^6$ nm/s, one has the
following unfolding pathways:
\begin{subequations}
\begin{equation}
G \rightarrow F \rightarrow (C,E,D) \rightarrow B \rightarrow A,
\label{pathways_v6_strand_eq}
\end{equation}
\begin{equation}
P_{AF} \rightarrow P_{BE} \rightarrow  (P_{FG}, P_{CF}) \rightarrow P_{CD}
\rightarrow P_{DE}
\rightarrow P_{AB}.
\label{pathways_v6_pair_eq}
\end{equation}
\end{subequations}
According to this scenario, the unfolding initiates from the C-terminal,
while the experiments \cite{Schwaiger_NSMB04} showed that
strands A and B unfold first.
For $v=2.6\times 10^4$ nm/s, Fig. \ref{cont_ext_v15_fig} gives
the following sequencing
\begin{subequations}
\begin{equation}
(A,B) \rightarrow (C,D,E) \rightarrow (F,G),
\label{pathways_v15_strand_eq}
\end{equation}
\begin{equation}
P_{AF} \rightarrow (P_{BE},P_{AB}) \rightarrow  P_{CF} \rightarrow 
(P_{CD},P_{DE},P_{FG}).
\label{pathways_v15_pair_eq}
\end{equation}
\end{subequations}
We obtain the very interesting result that at this low loading rate,
in agreement with the AFM experiments
\cite{Schwaiger_NSMB04}, the N-terminal detaches
from a protein first.

For both values of $v$, 
the first peak
corresponds to breaking of native contacts between
strands A and F (Fig. \ref{cont_ext_v6_fig}b and Fig. \ref{cont_ext_v15_fig}b).
However, the structure of unfolding intermediates, which correspond to this
peak, depends on $v$.
For $v=7.2\times 10^6$ nm/s (Fig. \ref{cont_ext_v6_fig}), at
$\Delta R \approx 1.5$ nm, native contacts between F and G are
broken and strand G has already
been unstructured (Fig. \ref{cont_ext_v6_fig}a). Therefore, for this
pulling speed, the intermediate consists of
six ordered strands A-F
(see Fig. \ref{snapshot_v6_v15_fig}a
for a typical snapshot).
In the $v=2.6\times 10^4$ nm/s case, just after the first peak, 
 none of strands
unfolds completely (Fig. \ref{cont_ext_v15_fig}a),
although
(A,F) and (B,E) contacts have been already broken (Fig. \ref{cont_ext_v15_fig}b).
Thus, the intermediate looks very different from the high $v$ case, as it has
all secondary structures partially structured
(see (Fig. \ref{snapshot_v6_v15_fig}b) for a typical snapshot).
Since the experiments \cite{Schwaiger_NSMB04}
showed that intermediate structures contain five ordered strands
C-G, intermediates predicted by simulations are more ordered than the
experimental ones. Even though,
our low loading rate Go simulations  provide the same pathways
as on the experiments. The difference between theory and experiments
in intermediate structures comes from
different locations of the first peak.
It remains unclear if this is a shortcoming of Go models or of
the experiments because it is hard to imagine that a $\beta$-protein like
 DDFLN4
displays
the first peak at such  a large extension $\Delta R \approx 12$ nm
\cite{Schwaiger_NSMB04}. The force-extension curve of
the titin domain I27, which has a similar native topology, for example, displays the
first peak at $\Delta R \approx 0.8$ nm \cite{Marszalek_Nature99}.
From this prospect, the theoretical result is more favorable.

The strong dependence of unfolding pathways on loading rates is 
also clearly seen from structures around the second peak.
In the $v=7.2\times 10^6$ nm/s case,
at $\Delta R \approx 11$ nm,
strands A and B remain structured, while other strands detach
from a protein core (Fig. \ref{cont_ext_v6_fig} and
Fig. \ref{snapshot_v6_v15_fig}c). This is entirely different from
the low loading case,
where A and B completely unfold
but F and G still survive (Fig. \ref{cont_ext_v15_fig} and
Fig. \ref{snapshot_v6_v15_fig}d).
The result, obtained for $v=2.6\times 10^4$ nm/s,
is in full agreement with the experiments \cite{Schwaiger_NSMB04}
that at $\Delta R \approx 12$ nm, A and B detached from the core. 

Note that the unfolding pathways given by Eq. \ref{pathways_v6_strand_eq},
\ref{pathways_v6_pair_eq},
\ref{pathways_v15_strand_eq}, and
\ref{pathways_v15_pair_eq}
are valid in the statistical sense. In all 50 trajectories studied
for $v=7.2\times 10^5$ nm/s, strands A and B always unfold last, and F and G
unfold first (Eq. \ref{pathways_v6_strand_eq}), while the sequencing of
unfolding events for C, D and E depends on individual trajectories.
At $v=2.6\times 10^4$ nm/s, most of trajectories follow
the pathway given by Eq. \ref{pathways_v15_strand_eq}, but
we have observed  a few unusual pathways, as it is illustrated in 
Fig. \ref{reentrance_FG_fig}. Having three peaks in 
the force-extension profile, 
the evolution of native contacts of
F and G display an atypical behavior. 
At $\Delta R \approx 7$ nm, these strands  fully unfold
(Fig. \ref{reentrance_FG_fig}c),
but they refold again at $\Delta R \approx 11$ nm (Fig. \ref{reentrance_FG_fig}b
and \ref{reentrance_FG_fig}d). Their final unfolding takes place
around $\Delta R \approx 16.5$ nm. As follows from
Fig. \ref{reentrance_FG_fig}b, the first peak in
Fig. \ref{reentrance_FG_fig}a corresponds to unfolding of G. Strands
A and B unfold after passing the second peak, while the third maximum occurs
due to unfolding of C-G , i.e. of a core part
shown in Fig. \ref{reentrance_FG_fig}d.

The dependence of unfolding pathways on $v$ is understandable.
If a protein is pulled very fast, the perturbation, caused
by the external force,  does not
have enough time to propagate to the fixed N-terminal before the C-terminal
unfolds. Therefore, at very high $v$, we have the pathway given by
Eq. \ref{pathways_v6_strand_eq}. In the opposite limit,
it does matter
what end is pulled as the external force
is uniformly felt along a chain. Then, a strand, which has
a weaker
link with the core, would unfold first.

\subsection{Computation of FEL parameters}

As mentioned above, at low loading rates, for some trajectories,
the force-extension curve
does not show two, but three peaks. However,
the percentage of such trajectories is rather
small,
we will neglect them and consider DDFLN4 as a three-state
protein.
Recently,
using dependencies of unfolding times on the 
constant external force and
the non-linear kinetic theory \cite{Dudko_PRL06},
we obtained distances $x_{u1} \approx x_{u2} \approx 13 \AA$
\cite{MSLi_JCP08}.
These values seem to be large for $\beta$-proteins like DDFLN4,
which are supposed to have smaller $x_u$ compared to $\alpha/\beta$- and
$\alpha$-ones \cite{MSLi_BJ07a}.
A clear difference between theory and experiments was also observed
for the unfolding barrier $\Delta G^{\ddagger}_1$.
 In order to see if one can improve our previous
results,
we will extract the FEL parameters by a different approach.
Namely, assuming that all FEL parameters of the three-state DDFLN4, 
including
the barrier between the second transition state and
the intermediate state $\Delta G^{\ddagger}_2$ (see 
Ref. \onlinecite{MSLi_JCP08} for the definition),
can be determined from dependencies of $f_{max1}$ and $f_{max2}$ on $v$,
we calculate them in the the Bell-Evans-Rirchie approximation
as well as beyond this approximation. 

{\bf Estimation of $x_{u1}$ and $x_{u2}$ in the Bell-Evans-Rirchie approximation}

In this approximation,
$x_{u1}$ and $x_{u2}$ are related to $v$, $f_{max1}$ and
$f_{max2}$ by the following equation \cite{Evans_BJ97}:
\begin{equation}
f_{maxi} \; = \; \frac{k_BT}{x_{ui} }
\ln \left[ \frac{vx_{ui}}{k_{ui}(0)k_BT}\right], i = 1,2,
\label{f_logV_eq}
\end{equation}
where $k_{ui}(0)$ is unfolding rates at zero external force.
In the low force regime ($v \lesssim 2\times 10^6$ nm/s), the 
dependence of $f_{max}$ on $v$ is logarithmic and  
$x_{u1}$ and $x_{u2}$ are defined by
slopes of linear fits in
Fig. \ref{fmax_Nfix_v_fig}. Their values are listed in Table 1.
The estimate of $x_{u2}$
agrees very well with the experimental \cite{Schwaiger_EMBO05}
as well as with the previous theoretical result \cite{MSLi_JCP08}.
The present value of
$x_{u1}$ agrees with the experiments better than the old one
\cite{MSLi_JCP08}.
Presumably, this is because it has been estimated by the same procedure
as in the experiments \cite{Schwaiger_EMBO05}.

It is important to note that
the logarithmic behavior is observed only at low enough $v$. At high
loading rates, the dependence of $f_{max}$ on $v$ becomes power-law.
This explains why all-atom simulations, performed 
at $v \sim  10^9$ nm/s for most of proteins, are not able to provide
reasonable estimations for $x_u$.

The another interesting question is if the peak at 
$\Delta R \approx 1.5$ nm disappears at loading rates used in the experiments
\cite{Schwaiger_EMBO05}. Assuming that the logarithmic dependence
in Fig. \ref{fmax_Nfix_v_fig} has the same slope at low $v$, we interpolate
our results to $v_{exp} = 200$ nm/s and obtain 
$f_{max1}(v_{exp}) \approx 40$ pN.
Thus, in the framework of the Go model, the existence of
the first peak is robust at experimental speeds.

{\bf Beyond the Bell-Evans-Rirchie approximation}

In the Bell-Evans-Rirchie approximation, one assumes that the location
of the transition state does not move under the action of an
external force. However,
our simulations for ubiquitin, for example, showed that it does move toward
the NS \cite{MSLi_BJ07}. 
Recently, assuming that $x_u$ depends on the external force
and using the Kramers theory, Dudko {\em et al.} have tried to
go beyond the Bell-Evans-Rirchie approximation. They proposed \cite{Dudko_PRL06}
the following formula for dependence of the unfolding force on
$x_u$ and $v$:
\begin{equation}
f_{max} \,  =  \frac{\Delta G^{\ddagger}}{\nu x_u} \left\{ 1- 
\left[\frac{k_BT}{\Delta G^{\ddagger}} \textrm{ln} \frac{k_BT k_u(0) e^{\Delta G^{\ddagger}/k_BT + \gamma}}{x_u v}\right]^{\nu} \right\}
\label{Dudko_eq}
\end{equation}
Here, $\Delta G^{\ddagger}$ is the unfolding barrier, $\nu = 1/2$ and 2/3
for the cusp \cite{Hummer_BJ03} and the
linear-cubic free energy surface \cite{Dudko_PNAS03}, respectively.
$\gamma \approx 0.577$ is the Euler-Mascheroni constant.
Note that
$\nu =1$ corresponds to the phenomenological
Bell theory (Eq. \ref{f_logV_eq}).
If $\nu \ne 1 $, then
Eq. \ref{Dudko_eq} can be used to estimate not only
$x_u$, but also $G^{\ddagger}$.
Since the fitting with $\nu = 1/2$ is valid in a wider force
 interval
compared to the $\nu = 2/3$ case, we
consider the former case only.
The region,
where the $\nu = 1/2$ fit works well, is expectantly wider  than that for
the Bell scenario (Fig. \ref{fmax_Nfix_v_fig}). 
From the nonlinear fitting (Eq. \ref{Dudko_eq}),
we obtain 
$x_{u1}=7.0 \AA\,$, and
$x_{u2} = 9.7 \AA\,$ which
are about twice as large as the Bell estimates (Table 1).
Using AFM data, Schlierf and Rief \cite{Schlierf_BJ06},
have shown that beyond Bell-Evans-Rirchie approximation
$x_u \approx 11 \AA\,$. This value is close to our estimate for $x_{u2}$.
However, a full comparison with experiments is not possible as
these authors did not consider $x_{u1}$ and $x_{u2}$ separately.
The present estimations of these quantities are
clearly lower than the previous one \cite{MSLi_JCP08} (Table 1).
The lower values
of $x_{u}$ would be more favorable because they are expected to
be not high for beta-rich proteins \cite{MSLi_BJ07a} like DDFLN4. 
Thus, beyond Bell-Evans-Rirchie approximation,
the method based on Eq. \ref{Dudko_eq} provides more reasonable
estimations for $x_{ui}$ compared to the method, where these
parameters are extracted
from unfolding rates \cite{MSLi_JCP08}. However,
in order to decide what method is better,
more experimental studies are required. 

The corresponding values for $G^{\ddagger}_1$, and $G^{\ddagger}_2$
are listed in Table 1.  
The experimental and previous theoretical
results \cite{MSLi_JCP08} are also shown for comparison.
The present estimates for both barriers agree with the 
experimental data, while  
the previous theoretical value of $\Delta G^{\ddagger}_1$ 
fits to experiments worse than
the current one.

\subsection{Thermal unfolding pathways}

In order to see if the thermal unfolding pathways are different
from the mechanical ones, we performed zero-force simulations
at $T=410$ K.  The progress variable $\delta$ is used
as a reaction coordinate to monitor pathways (see {\em Materials and Methods}).
From Fig. \ref{ther_unfold_pathways_snap_fig}, we have  the following sequencing
for strands and their pairs:
\begin{subequations}
\begin{equation}
G \rightarrow (B, C, E)  \rightarrow (A, F, D),
\label{thermal_pathway_str_eq}
\end{equation}
\begin{equation}
P_{AF}  \rightarrow P_{BE}  \rightarrow (P_{CD}, P_{CF})  \rightarrow 
(P{AB}, P_{FG}, P_{DE}).
\label{thermal_pathway_pair_eq}
\end{equation}
\end{subequations}
It should be noted that these pathways are just 
major ones as other pathways
are also possible. The pathway given by Eq. \ref{thermal_pathway_pair_eq},
e.g.,  occurs in 35\% of events.
About 20\% of trajectories follow
$P_{AF}  \rightarrow P_{CF}  \rightarrow P_{BE} \rightarrow
(P_{CD},P{AB}, P_{FG}, P_{DE})$ scenario. We have also observed the sequencing
 $P_{AF}  \rightarrow P_{BE} \rightarrow
(P_{CF},P{AB}, P_{FG}, P_{DE}) \rightarrow P_{CD}$, and
$P_{BE}  \rightarrow P_{AF} \rightarrow
(P_{CD},P{CF}, P_{AB}, P_{FG},P_{DE})$ in 12\% and 10\% of runs, respectively.
Thus,
due to strong thermal fluctuations,
thermal unfolding pathways are more diverse compared to mechanical ones.
From Eqs. \ref{pathways_v6_strand_eq}, \ref{pathways_v6_pair_eq},
\ref{pathways_v15_strand_eq}, \ref{pathways_v15_pair_eq},
\ref{thermal_pathway_str_eq}, and \ref{thermal_pathway_pair_eq}, it is clear that thermal unfolding
pathways of DDFLN4 are different from the mechanical pathways.
This is also
illustrated in 
Fig. \ref{ther_unfold_pathways_snap_fig}c.
As in the mechanical case
(Fig. \ref{snapshot_v6_v15_fig}a and \ref{snapshot_v6_v15_fig}b),
 the contact between A and F is broken,
but the molecule is much
less compact at the same end-to-end distance. 
Although 7 contacts ($\approx 64$\%) between strands
F and G remain survive, all contacts of pairs
$P_{AF}, P_{BE}$ and $P_{CD}$ are already broken.

The difference between mechanical and thermal unfolding pathways
is attributed to the fact
that thermal fluctuations have a global effect on the biomolecule,
while the force acts only on its termini. Such a difference was also observed
for other proteins like I27 \cite{Paci_PNAS00} and ubiquitin
\cite{MSLi_BJ07,Mitternacht_Proteins06}.
We have also studied folding pathways of DDFLN4 at $T=285$ K. It turns
out that they are reverse of the thermal unfolding pathways given
by Eqs. \ref{thermal_pathway_str_eq} and  \ref{thermal_pathway_pair_eq}.
It would be interesting to test our prediction on thermal folding/unfolding
of this domain experimentally.

\noindent{\bf Conclusions}

The key result of this paper is that
mechnanical unfolding pathways of DDFLN4 depend on loading rates.
At large $v$ the C-terminal unfolds first, but the N-terminal
unfolds at low $v \sim 10^4$ nm/s. The agreement with the
experiments \cite{Schwaiger_NSMB04}
is obtained only
in low loading rate simulations.
The dependence of mechanical unfolding pathways on the loading rates
was also observed for I27 (M.S. Li, unpublished). On the other hand,
the previous studies \cite{Irback_PNAS05,MSLi_BJ07} showed that mechanical unfolding pathways
of the two-state ubiquitin do not depend on the force strength.
Since DDFLN4 and I27 are three-state
proteins, one may think that
the unfolding pathway change with variation of the pulling
speed,
is universal 
for proteins that unfold via intermediates.
A more comprehensive study is needed to verify this 
interesting issue.

Dependencies of unfolding forces on pulling speeds have been widely used
to probe FEL of two-state proteins \cite{Best_PNAS02}.
However, to our best knowledge,
here we have made a first attempt to apply this approach
to extract not only
$x_{ui}$, but also 
$\Delta G^{\ddagger}_i$ ($i= 1,$ and 2) for a three-state protein.
This allows us to improve our previous results \cite{MSLi_JCP08}.
More importantly, a better agreement with the experimental data
\cite{Schwaiger_EMBO05,Schlierf_BJ06} suggests that this method
is also applicable to
other multi-state  biomolecules.
Our study clearly shows that the low loading
rate regime, where FEL parameters can be estimated, occurs at 
$ v \leq 10^6$ nm/s which are about two-three orders of magnitude
lower than those used in all-atom simulations.
Therefore, at present, deciphering unfolding FEL of long proteins by
all-atom simulations with explicit water is computationally prohibited.
From this point of view, coarse-grained models are of great help.

We predict the existence of a peak at $\Delta R \sim 1.5$ nm even
at pulling speeds used in now a day experimental setups.
One of possible reasons of why the experiments did not detect this
maximum is realted to a strong linker effect as
a single DDFLN4 domain is sandwiched between Ig domains
I27-30 and domains I31-34 from titin \cite{Schwaiger_NSMB04}.
Therefore, our result would stimulate
new experiments on mechanical properties of this protein.
Capturing the experimentally observed peak at $\Delta R \sim 22$ nm
remains a challenge to theory.

{\em
Mechanical unfolding pathways of DDFLN4 and other proteins
\cite{Paci_PNAS00,MSLi_BJ07,Mitternacht_Proteins06} are different from
thermal ones. In accord with
a common belief \cite{Daggett_TBS03}, 
thermal unfolding pathways of these
proteins were shown to be reverse of folding pathways.
Therefore,
their folding mechanisms can not be gained from mechanical studies.
Recently, using the all-atom simulations with implicit solvent
\cite{Irback_JCC06},
it has been found that 
a 49-residue C-terminal of TOP7 (residues 2-50 of 2GJH.pdb)
 folds via a non-trivial caching mechanism
\cite{Mohanty_PNAS08} and its thermal unfolding pathways are not 
reverse of the folding ones \cite{Mohanty_JPCB08}.
Can the folding mechanism of
this fragment be deduced from mechanical unfolding simulations and experiments?
A detailed study of this interesting question is in progress but our
preliminary simulation results show that folding pathways
may be inferred from the mechanical ones.
}

The work was 
supported by
the Ministry of Science and Informatics in Poland
(grant No 202-204-234).
MK is very grateful to the Polish committee for UNESCO
for the financial support.


\newpage

\begin{center}
\begin{tabular}{lll|lllr}
& \multicolumn{2}{c|}{Bell approximation} &\multicolumn{4}{c}{Beyond Bell approximation}\\ \cline{2-7}
& \; $x_{u1}(\AA)$ \;& \; $x_{u2}(\AA)$ \; & \; $x_{u1}(\AA)$ \;& \; $x_{u2}(\AA)$ \; & \; $\Delta G^{\ddagger}_1/k_BT \;$ &  \; $\Delta G^{\ddagger}_2/k_BT \; $ \\
\hline
Theory \cite{MSLi_JCP08}&\; 6.3 $\pm$ 0.2 \; & \; 5.1 $\pm$ 0.2 \;&\; 13.1 \; &\; 12.6 \; & \; 25.8 \; & \; 18.7 \; \\
Theory (this work)&\; 3.2 $\pm$ 0.2 \; & \; 5.5 $\pm$ 0.2 \;&\; 7.0 \;& \; 9.7 \; & \; 19.9 \; & \; \; 20.9 \; \\
Exp. \cite{Schwaiger_EMBO05,Schlierf_BJ06} & \; 4.0 $\pm 0.4$ \; & \; 5.3 $\pm$ 0.4 \; & & &\; 17.4 \; & \; 17.2 \;\\
\hline
 \end{tabular}
\end{center}

\vspace{0.5cm}
Table 1. Parameters $x_{u1}$, and $x_{u2}$ were obtained in the
Bell and beyond-Bell approximation. Theoretical values of the unfolding
barriers were extracted from the microscopic theory of Dudko {\em et al}
(Eq. \ref{Dudko_eq}) 
with $\nu = 1/2$. The experimental estimates were taken from
Ref. \onlinecite{MSLi_JCP08}.

\newpage

\centerline{\Large \bf Figure Captions}

\vskip 5mm

\noindent {\bf FIGURE 1.} (a) Native state conformation of 
DDFLN4 taken from the PDB
(PDB ID: 1ksr). There are seven $\beta$-strands: A (6-9), B (22-28),
C (43-48), D (57-59), E (64-69), F (75-83), and
G (94-97).
In the native state there are 15, 39, 23, 10, 27, 49, and 20 native contacts
formed by strands A, B, C, D, E, F, and G with
the rest of the protein, respectively.
The end-to-end distance in the native state $R_{NS}=40.2$ \AA.
(b) There are 7 pairs of strands, which have the nonzero number
 of mutual native contacts
in the native state. These pairs are  P$_{\textrm{AB}}$, 
P$_{\textrm{AF}}$, P$_{\textrm{BE}}$,
P$_{\textrm{CD}}$, P$_{\textrm{CF}}$, P$_{\textrm{DE}}$, and P$_{\textrm{FG}}$.
The number of native contacts between them
 are 11, 1, 13, 2, 16, 8, and 11,
respectively.
 \vskip 5 mm

\noindent {\bf FIGURE 2.}Typical force-extension curves for
 $v =7.2\times 10^6$ nm/s (a), $6.4\times 10^5$ nm/s (b),
$5.8\times 10^4$ nm/s (c), and 2.6$\times 10^4$  mn/s (d). The arrow
in (c) and (d)
roughly refers to locations of additional peaks for two trajectories (red and green).

\vskip 5 mm

\noindent {\bf FIGURE 3.} Distributions of positions of $f_{max1}$ and
$f_{max2}$ for $v =7.2\times 10^6$ (solid), $ 6.4\times 10^5$ (dashed) , $5.8\times 10^4$ (dotted) and 2.6$\times 10^4$  mn/s (dashed-dotted).

\vskip 5 mm

\noindent {\bf FIGURE 4.} (a) Dependence of averaged
fractions of native contacts
formed by seven strands on $\Delta R$ for $v = 7.2\times 10^6$ nm/s.
(b) The same as in (a) but for pairs of strands.
Arrows refer to the positions of peaks.
Results were averaged over 50 trajectories.

\vskip 5 mm

\begin{figure}
\epsfxsize=5in
\vspace{0.2in}
\centerline{\epsffile{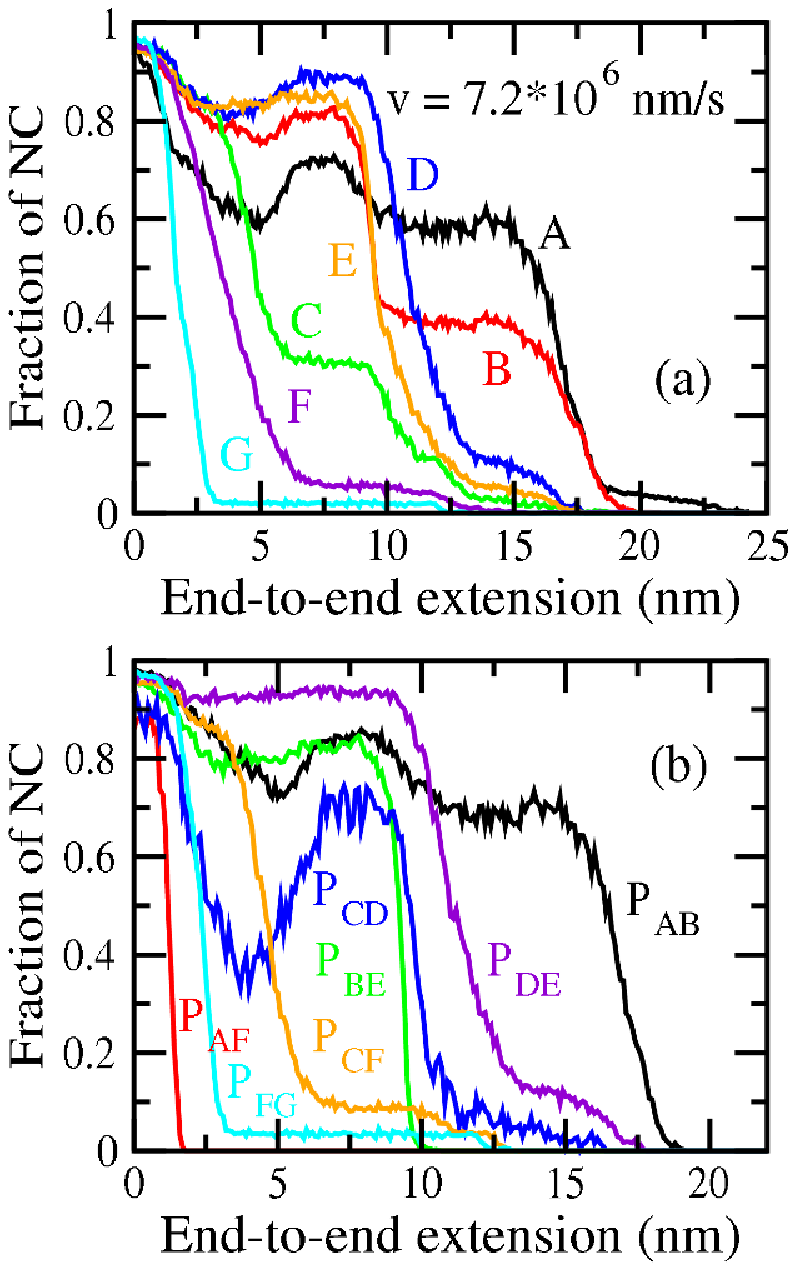}}
\caption{}
\label{cont_ext_v6_fig}
\end{figure}

\begin{figure}
\epsfxsize=5in
\vspace{0.2in}
\centerline{\epsffile{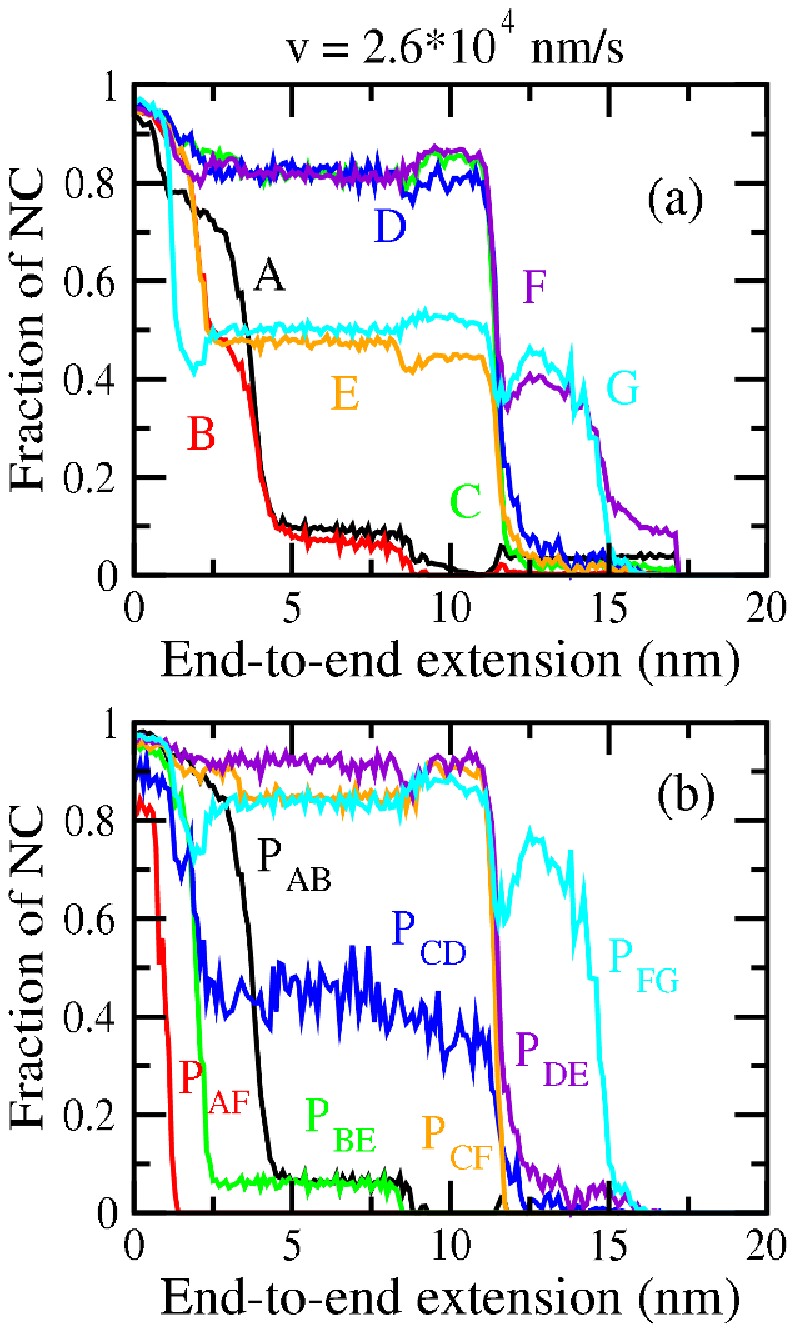}}
\caption{}
\label{cont_ext_v15_fig}
\end{figure}

\begin{figure}
\epsfxsize=6.3in
\vspace{0.2in}
\centerline{\epsffile{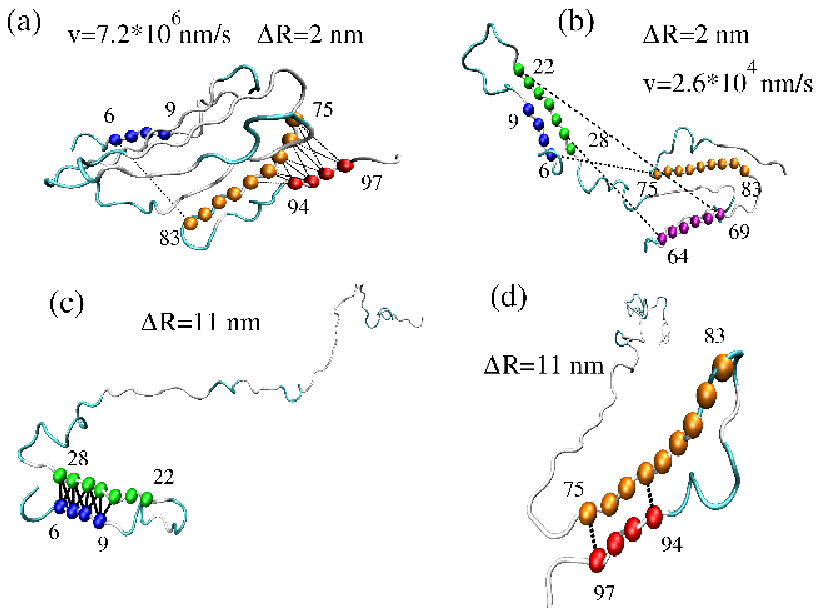}}
\caption{}
\label{snapshot_v6_v15_fig}
\end{figure}

\begin{figure}
\epsfxsize=6.3in
\vspace{0.2in}
\centerline{\epsffile{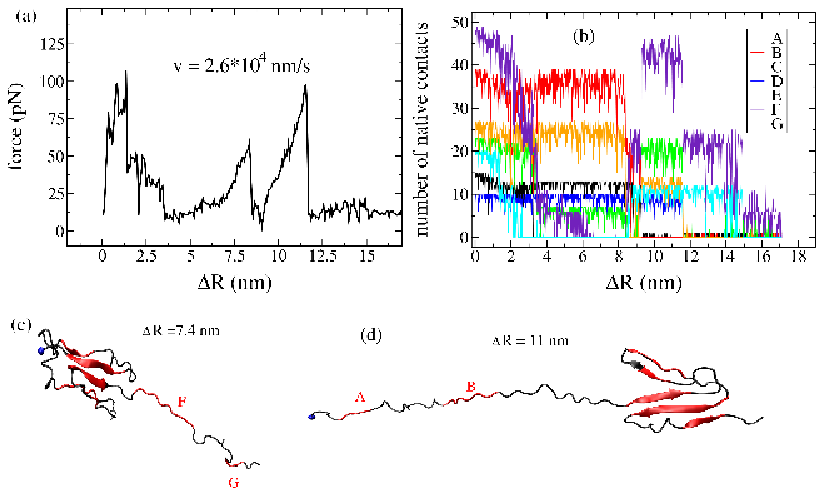}}
\caption{
\label{reentrance_FG_fig}}
\end{figure}

\begin{figure}
\epsfxsize=6.3in
\vspace{0.2in}
\centerline{\epsffile{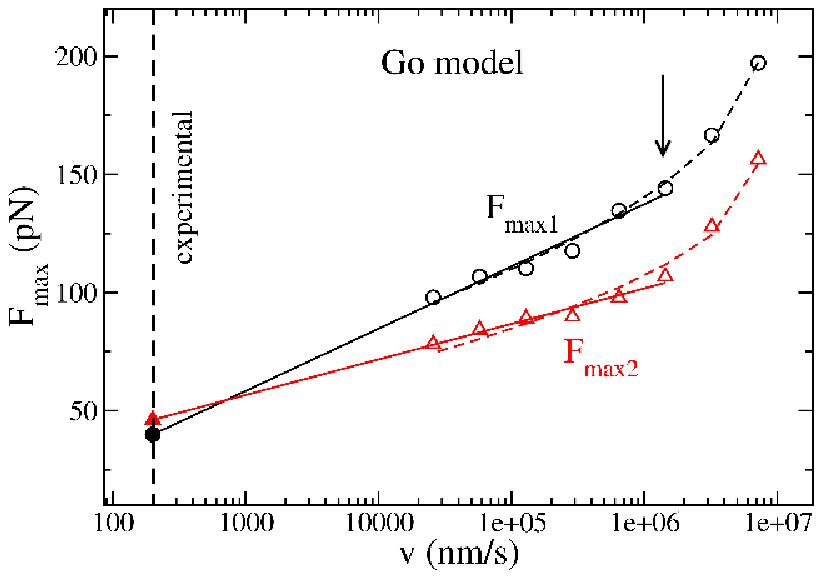}}
\caption{}
\label{fmax_Nfix_v_fig}
\end{figure}

\begin{figure}
\epsfxsize=6.3in
\vspace{0.2in}
\centerline{\epsffile{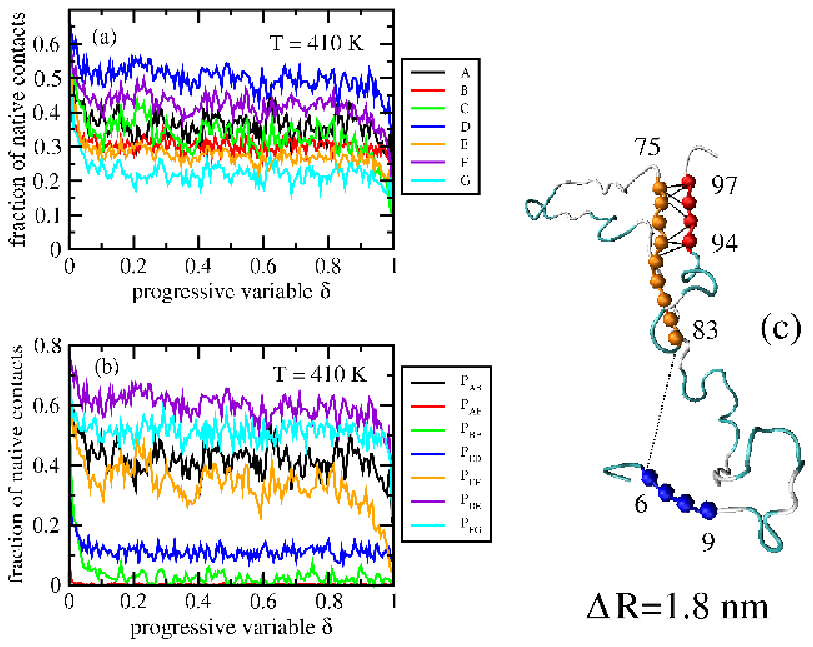}}
\caption{}
\label{ther_unfold_pathways_snap_fig}
\end{figure}

\noindent {\bf FIGURE 5.} The same as in Fig. \ref{cont_ext_v6_fig} but
for $v=2.6\times 10^4$ nm/s. 
Results were averaged over 50 trajectories.

\vskip 5 mm

\noindent {\bf FIGURE 6.} (a) Typical snapshot obtained at $\Delta R = 2$ nm
and $v= 7.2\times 10^6$ nm/s. A single contact between strand A (blue spheres)
and strand F (orange) was broken (dotted lines). Native contacts between F and G (red) are also
broken and G completely unfolds. (b) The same as in (a) but
for $v=2.6\times 10^4$ nm/s. Native contacts between A and F and between
B and E are broken (dotted lines), but all strands are remain partially structured.
(c) Typical snapshot obtained at $\Delta R = 11$ nm
and $v= 7.2\times 10^6$ nm/s. Native contacts between pairs are broken except
those between strands A and B. 
All 11 unbroken contacts are marked by solid lines.  Strands A and B
do not unfold yet.
(d) The same as in (c) but for $v=2.6\times 10^4$ nm/s.
Two from 11 native contacts between F and G are broken (dotted lines).
Contacts between other pairs are already broken, but F and G remain structured.

\vskip 5 mm

\noindent {\bf FIGURE 7.} (a) Force-extension curve for
an anomalous unfolding  pathway at $v = 2.6\times 10^4$ nm/s. (b)
Dependence of fractions of native contacts of seven strands on $\Delta R$.
Snapshot at $\Delta R = 7.4$ nm (c) and $\Delta R = 11$ nm (d).

\vskip 5 mm

\noindent {\bf FIGURE 8.} Dependence of $f_{max1}$ (open circles) 
and $f_{max2}$ (open squares) on $v$.
The values of these peaks were
obtained as averages over all trajectories.
The arrow separates the low pulling speed regime from
the high one.
Straight lines are fits to the Bell-Evans-Rirchie equation
($y = -20.33 + 11.424ln(x)$ and $y= 11.54 + 6.528ln(x)$ for $F_{max1}$
and $F_{max2}$, respectively). Here $f_{max}$ and $v$ are measured in pN
and nm/s, respectively.
From these fits we obtain
$x_{u1}=3.2 \AA\,$ and $x_{u2}=5.5 \AA$.
The solid circle and triangle correspond to $f_{max1} \approx 40$ pN
and $f_{max2} \approx 46$ pN,
obtained by interpolation of linear fits to the experimental
value $v = 200$ nm/s. 
Fitting to the nonlinear microscopic theory (dashed lines) gives
$x_{u1}=7.0 \AA\, , \Delta G^{\ddagger}_1 = 19.9 k_BT,
x_{u2} = 9.7 \AA\,$, and $\Delta G^{\ddagger}_2 = 20.9 k_BT$.

\vskip 5 mm

\noindent {\bf FIGURE 9.} Thermal unfolding pathways. (a) Dependence of
native contact fractions of
seven strands on the progress variable $\delta$ at $T=410$ K.
(b) The same as in (a) but for seven strand pairs.
(c) A typical snapshot at
$\Delta R \approx 1.8$ nm. The contact between
strands A and F is broken (dotted lines)
 but 7 contacts between strands S6 and S7
(solid lines) still
survive.

\clearpage

\clearpage

\clearpage

\clearpage

\clearpage

\end{document}